\RequirePackage[hyphens]{url}
\documentclass[aps,prb,twocolumn,showpacs,bibnotes,reprint,superscriptaddress]{revtex4-1}

\pdfoutput=1

\usepackage{hyperref,url}
\usepackage{tikz}
\usetikzlibrary{positioning,arrows}
\usepackage{paralist}
\usepackage[compact]{titlesec}
\titlespacing{\section}{10pt}{2.5ex}{1.5ex}

\hypersetup{hyperindex,breaklinks,colorlinks=true,linkcolor=blue,citecolor=blue,urlcolor=blue,filecolor=blue}

\begin{document}

\title{Open-source development experiences in scientific software: the HANDE quantum Monte Carlo project}

\author{J.~S.~Spencer}
\thanks{\href{mailto:j.spencer@imperial.ac.uk}{j.spencer@imperial.ac.uk}, \href{mailto:ajwt3@cam.ac.uk}{ajwt3@cam.ac.uk}}
\affiliation{Department of Materials, Imperial College London, Exhibition Road, London, SW7 2AZ, United Kingdom}
\affiliation{Department of Physics, Imperial College London, Exhibition Road, London, SW7 2AZ, United Kingdom}
\author{N.~S.~Blunt}
\affiliation{University Chemical Laboratory, Lensfield Road, Cambridge, CB2 1EW, United Kingdom}
\author{W.~A.~Vigor}
\affiliation{Department of Chemistry, Imperial College London, Exhibition Road, London, SW7 2AZ, United Kingdom}
\author{Fionn~D.~Malone}
\affiliation{Department of Physics, Imperial College London, Exhibition Road, London, SW7 2AZ, United Kingdom}
\author{W.~M.~C.~Foulkes}
\affiliation{Department of Physics, Imperial College London, Exhibition Road, London, SW7 2AZ, United Kingdom}
\author{James~J.~Shepherd}
\affiliation{Department of Chemistry, Rice University, Houston, TX 77005-1892, USA}
\author{A.~J.~W.~Thom}
\thanks{\href{mailto:j.spencer@imperial.ac.uk}{j.spencer@imperial.ac.uk}, \href{mailto:ajwt3@cam.ac.uk}{ajwt3@cam.ac.uk}}
\affiliation{University Chemical Laboratory, Lensfield Road, Cambridge, CB2 1EW, United Kingdom}

\begin{abstract}
    The HANDE quantum Monte Carlo project offers accessible stochastic algorithms for general use for scientists in the field of quantum chemistry.
    HANDE is an ambitious and general high-performance code developed by a geographically-dispersed team with a variety of backgrounds in computational science.
    In the course of preparing a public, open-source release, we have taken this opportunity to step back and look at what we have done and what we hope to do in the future.  We pay particular attention to development processes, the approach taken to train students joining the project, and how a flat hierarchical structure aids communication.
\end{abstract}

\date{\today}
\maketitle

The Highly Accurate N-DEterminant (HANDE) quantum Monte Carlo project\cite{hande} began life as an experiment by one of us (JSS) to explore the (then recent) development in quantum chemistry: the full configuration interaction quantum Monte Carlo (FCIQMC) method~\cite{booth:054106}.
FCIQMC can be viewed simply as a stochastic approach to the power method; it allows the calculation of exact ground state energies of quantum systems with Hilbert spaces orders of magnitude larger than accessible via even state-of-the-art deterministic algorithms.
Initially only the Hubbard model was implemented, but HANDE now handles a range of model and chemical systems.
At the same time HANDE has become an efficient and highly parallel implementation of FCIQMC and related methods\cite{cleland_initiator_2010,petruzielo_semistochastic_2012}, capable of scaling to several thousand cores.
We have also provided deeper understanding of the FCIQMC method\cite{Spencer2012,Kolodrubetz2013,Shepherd2014}, extended HANDE to include the canonical implementation of the stochastic coupled cluster approach\cite{thom_ccmc_2010} and developed new methods within the field\cite{blunt_density-matrix_2014}.
The driving-force for this transformation, from a toy code to a professional software package, has been the team of contributors split between three universities working together in a sustainable and robust process.
We are very proud of the variety of our developers, who represent several different areas of science and range from undergraduates to professors.
Indeed, we have had exceptional success with undergraduate research projects, which is remarkable given that most start with no or little experience in parallel computing and in quantum chemistry---a notable example is the development of a novel Monte Carlo method by two undergraduate students\cite{blunt_density-matrix_2014}.

The unexpected and organic growth has provided its challenges.  How to transition into a community-owned code from the initial gatekeeper model we stumbled into?  How to develop and support new contributors to the project?  In some cases we planned ahead; in others we reached a consensus through iterative experimentation.  Indeed, we have found flexibility and willingness to adapt to be of vital importance.

In this contribution we first describe the choices we made in an effort to write a sustainable, portable library, the approach we have settled on for development and the benefits we have subsequently obtained.  We then discuss how we have trained students to be successful and valuable members of the development team and our future plans for the HANDE project before offering our conclusions and suggestions to the wider computational science community.

\section*{HANDE OVERVIEW}

HANDE is a small, but growing, project with half a dozen active developers at any one time.  Most users are also developers but the user community is growing through active collaborations.  The code base contains approximately 20000 lines of Fortran 2003, plus a smaller amount of C and several thousand lines of comments and is parallelised using MPI and OpenMP.  HANDE is available as a source distribution via the project website\cite{hande} and github\cite{hande_github}.  The distribution also contains a substantial amount of documentation, including compilation and usage instructions, and tutorials as well as python modules for data analysis.  HANDE is developed on Linux, Mac OS X and Windows though, due to the nature of supercomputers, production calculations on high performance computer facilities are universally performed on Linux.

\section*{A DEVELOPMENT MODEL}

We view ourselves as scientists \emph{and} programmers (though our funding agencies might not agree!) and believe both roles are vital.  As programmers, a maintainable and efficient code is our main goal.  As scientists, we wish to rapidly address the questions posed in our research.  These positions are not, however, contradictory: rather we have found the programmers' goal also minimizes delays in making scientific progress once spread over a number of consecutive projects.  In other words, poor design and development choices eventually hinder us.  Here we detail some of the choices we have made and their consequences. We note that the comments we have to make are surprisingly general; an in-depth knowledge of the algorithm is not necessary to appreciate what we are discussing.  We are, however, aided by FCIQMC and related methods being simple and composed of only a few distinct data flows.  In particular, the memory demands are dominated by the representation of the eigenvector and the computational cost per iteration by the tight loop in which the eigenvector is stochastically evolved.

{\bf Coding conventions}--- We have taken care to maintain consistency in coding conventions throughout.
This begins with a common, ordered commenting style\cite{hande_github_ccmc}; this visual cue helps developers become immediately aware of the existence of code norms and leads to it being easier to maintain wide-spread adoption of the other features below.
Apart from making it far easier and more pleasant to read and understand code, such conventions serve as a guide to those with little prior experience programming and help prevent code from being rushed.
We ensure that the functionality and inputs and outputs of all procedure interfaces are documented; this can then be extracted using tools such as sphinx\cite{sphinx} and makes comprehension whilst navigating code (e.g.\ using ctags\cite{ctags}) far faster.
We further advocate the use of extensive commenting to provide both an overview of the theory and the choices that lie behind an implementation: indeed, in the more theoretically challenging parts of HANDE, the amount of comments rivals or exceeds the actual amount of code.
Such cases can be viewed as an example of literate programming and may include theoretical overviews (which, for research software, are frequently not yet available in the literature), a discussion on implementation choices, benchmarks, examples and so on.
These serve both as documentation and as extremely helpful material from which new members of the development team can learn about details which may be inappropriate for traditional papers.

{\bf Pure functions}--- A growing trend in HANDE development, which has been successful, is a move towards the use of pure functions, which (along with other functional programming approaches) have been demonstrated to have compelling advantages\cite{bacus1978,orchard2014,ray2014}. The results of pure functions depend only upon the input argument values and have no side effects on any part of the code outside the function. As such, pure functions cannot depend on any global data. We have found that functions which depend heavily on global data have many subtle interactions and assumptions, such that changing one part of the code can unexpectedly alter other parts. This problem becomes worse as the size of a program grows. In contrast, one can be confident that changes outside a pure function can never alter its results for the same set of inputs. Beyond this, code written in a pure style is more reusable (both within the code and in separate projects) and easier to test.  Whilst writing code in a pure style can initially take longer, we are finding that it saves significant time and effort in the long run and makes implementing new functionality far easier. We have utilized this for threaded parallelism and alternate implementations.

{\bf Factorisation}--- Open source software provides a huge advantage to our developers; they are encouraged to extract code which could be reused in other projects to contribute to the community.
This approach to factorisation forces developers to plan and separate functionally and logically independent code, improving the quality and sustainability of the code.  Conversely we benefit from similar efforts in the broader community and can use state-of-art portable libraries to minimise time-to-science and avoid duplication of effort.
For example, we use HDF5 for checkpoint files~\cite{hdf5}, dSFMT for random numbers~\cite{dSFMT} and the python scientific stack (especially numpy~\cite{numpy}, pandas~\cite{pandas} and matplotlib~\cite{matplotlib}) for data analysis.
In return, our contributions include Fortran interfaces to libraries~\cite{dsfmtinterface}, a test framework~\cite{testcode} (see below) and a python library for removing serial correlations in Monte Carlo data~\cite{pyblock}.
We find such efforts are a way of broadening impact of our development work far beyond the immediate stochastic quantum chemistry community.  Encouragingly, we have also received contributions to these libraries from outside of our team.  Making the code publicly accessible via distributed version control (e.g.\ on github) is key to reducing the barrier to entry.

Despite the above, a large number of dependencies is undesirable from a usability viewpoint: requiring the user to manually compile several packages before using our program hinders experimentation and porting to new platforms.  We try to overcome this in two ways: small libraries with permissive licenses can be included in the source distribution and non-core features which depend upon larger libraries can be disabled at compile-time.

{\bf Pull requests and code review}--- In the last year we have moved to a system of pull requests based upon the git flow model\cite{git-flow}. In this system, any contributions to HANDE must be made on a branch (using our version control system of choice, git) and a review of the branch performed (by at least one other contributor) before it may be merged into master (see Fig.~1).  Code review can easily be performed using (e.g.) github's inline commenting or, our preferred tool, watson\cite{watson}. Code review is deliberately light weight and allows for rapid peer feedback about the approach used, problems in the design and consistency in code style.
In particular, the process typically includes validation and verification, of the code, documentation and (crucially) any new theoretical work underlying it.
We have found that this process greatly reduces bugs and rushed code from ending up in the master, which is designated to be sufficiently stable for production calculations.  Already we have seen substantial improvements in the flexibility, sustainability and maintainability of the code.
It also gives contributors an understanding of parts of the codebase that they may not otherwise know much about. Even those who do not perform a review in detail gain knowledge of the various projects being worked on.
The social impact of this is interesting: we find code review to be an excellent way of flattening the academic hierarchical structure.  In particular, we note that the levels of expertise in scientific and computational domains are often not aligned and the more `junior' members of a research team are often the ones doing the most software development and hence their reviews of contributions from more `senior' members can be the most enlightening.

One aspect deserves special consideration: not all development work is evolutionary; some must be revolutionary.  This kind of development work is frequently long running and handling both the review and merging (often into a very different codebase after months of parallel development) is painful.  We have found that regular peer review of intermediate work and occasional rebasing of such branches against the current development version of the code goes a long way to mitigating such issues.

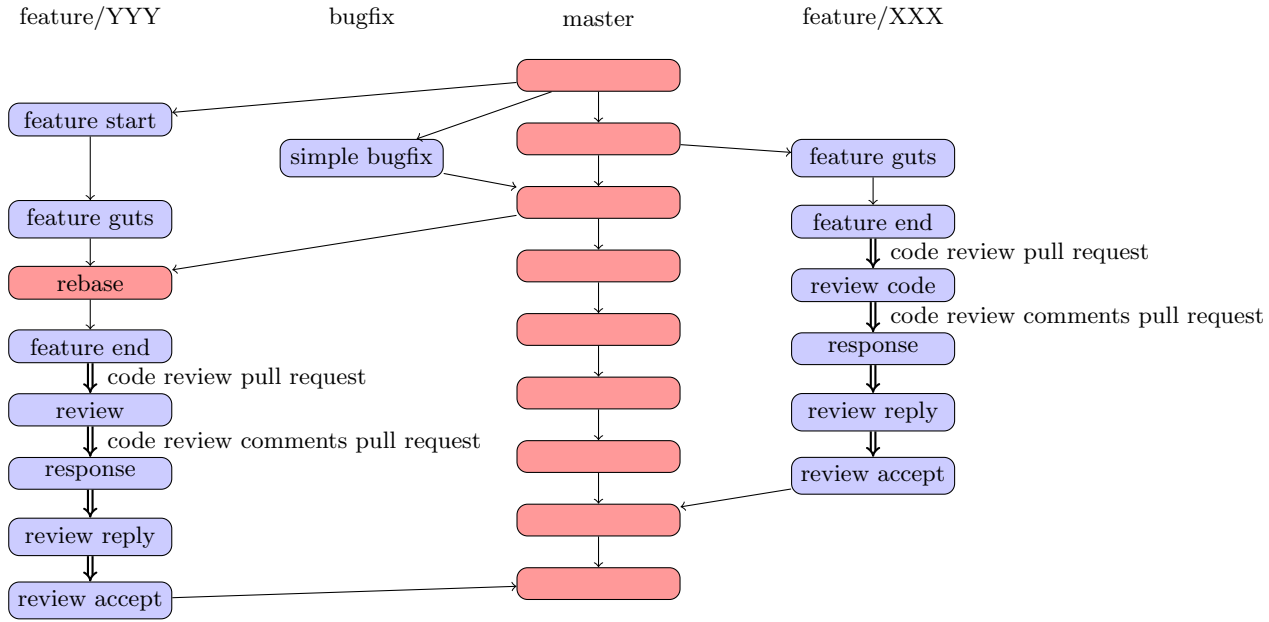
\begin{figure*}
\vspace*{-0.5cm}
\tikzstyle{simple} = [rectangle, draw, fill=blue!20, 
    text width=6em, text centered, rounded corners, minimum height=1.3em]
\tikzstyle{merge} = [rectangle, draw, fill=red!40, 
    text width=6em, text centered, rounded corners, minimum height=1.3em]
\tikzstyle{email} = [-implies,double equal sign distance,thick]
\begin{tikzpicture}[node distance=2.6em, auto]
\node(master) {master};
\node[merge, below=1em of master] (master0) {};
\node[merge, below of=master0] (master1) {};
\draw [->] (master0) -- (master1) {};
\node[merge, below of=master1] (master2) {};
\draw [->] (master1) -- (master2) {};
\node[merge, below of=master2] (master3) {};
\draw [->] (master2) -- (master3) {};
\node[merge, below of=master3] (master4) {};
\draw [->] (master3) -- (master4) {};
\node[merge, below of=master4] (master5) {};
\draw [->] (master4) -- (master5) {};
\node[merge, below of=master5] (master6) {};
\draw [->] (master5) -- (master6) {};
\node[merge, below of=master6] (master7) {};
\draw [->] (master6) -- (master7) {};
\node[merge, below of=master7] (master8) {};
\draw [->] (master7) -- (master8) {};

\node[right=2cm of master] (feature) {feature/XXX};
\node[simple, below=of feature |- master0] (feature1) {feature guts};
\draw [->] (master1) -- (feature1) {};
\node[simple, below of=feature1] (feature2) {feature end};
\draw [->] (feature1) -- (feature2) {};
\node[simple, below of=feature2] (feature3) {review code};
\draw [email,right] (feature2) --  node {\ code review pull request} (feature3);
\node[simple, below of=feature3] (feature4) {response};
\draw [email,right] (feature3) --  node {\ code review comments pull request}(feature4);
\node[simple, below of=feature4] (feature5) {review reply};
\draw [email] (feature4) -- (feature5) {};
\node[simple, below of=feature5] (feature6) {review accept};
\draw [email] (feature5) -- (feature6) {};
\draw [->] (feature6) -- (master7) {};

\node[left=2cm of master] (bugfix) {bugfix};
\node[simple, below=of bugfix |- master0] (bugfix1) {simple bugfix};
\draw [->] (master0) -- (bugfix1) {};
\draw [->] (bugfix1) -- (master2) {};

\node[left=2cm of bugfix] (featurey) {feature/YYY};
\node[simple, below=of featurey ] (featurey1) {feature start};
\draw [->] (master0) -- (featurey1) {};
\node[simple, below=of featurey1] (featurey1a) {feature guts};
\draw [->] (featurey1) -- (featurey1a) {};
\node[merge, below of=featurey1a] (featurey2) {rebase};
\draw [->] (featurey1a) -- (featurey2) {};
\draw [->] (master2) -- (featurey2) {};
\node[simple, below of=featurey2] (featurey3) {feature end};
\draw [->] (featurey2) -- (featurey3) {};
\node[simple, below of=featurey3] (featurey4) {review};
\draw [email,right] (featurey3) --  node {\ code review pull request} (featurey4);
\node[simple, below of=featurey4] (featurey5) {response};
\draw [email,right] (featurey4) --  node {\ code review comments pull request}(featurey5);
\node[simple, below of=featurey5] (featurey6) {review reply};
\draw [email] (featurey5) -- (featurey6) {};
\node[simple, below of=featurey6] (featurey7) {review accept};
\draw [email] (featurey6) -- (featurey7) {};
\draw [->] (featurey7) -- (master8) {};
\end{tikzpicture}
\caption{Git workflow.  Blue indicates a simple commit, and red a merge commit.  As all changes are made in a branch and merged to master, all master commits are merges and undergo automated integration and regression testing. Not all branches are shown for simplicity. Double arrows are accompanied by an email to the developer list.}
\vspace*{-0.25cm}
\end{figure*}

{\bf Regression testing}--- Scientific codes produce quantitative results that, in principle, should be extremely simple to test against when the code changes.
When differences happen to indicate a bug, these can be tracked down between a relatively small number of commits using a bisection method.
Whilst unit tests are valuable, we have found that regression tests are easier to retrofit to existing code bases and are good at capturing problems in the interfaces between procedures or changes compared to existing answers.
This type of regression testing is relatively straight-forward to undertake.  Apart from data extraction from output files, regression testing involves a generic set of tasks.
One of us (JSS) maintains an open source portable tool for just such a purpose~\cite{testcode}, which has attracted use in the wider electronic structure community.
Running the tests can be automated (e.g. to check every commit, every pull request, given time intervals) using tools such as jenkins, travis-ci or buildbot, which is currently used in the HANDE project, as is performed by many other projects (e.g.~\onlinecite{gaston2014,CASTEP}).
The design of tests themselves is a non-trivial challenge, and should not be underestimated.
A test should check a broad sweep of functionality, but when there are many input parameters (and variably sparse matrices) it is impossible to check every combination, though tools such as gcov are invaluable in discovering the fraction of the code covered by a set of tests.
HANDE contains over 160 tests which cover over 85\% of the code base (excluding external libraries) and increasing this is an ongoing effort.
Moreover, because the software is designed for high-performance computing and contains Monte Carlo algorithms, it can be hard to reliably review this functionality, especially for bugs which are only revealed when run on thousands of processors.
Where possible, therefore, new conceptual developments are checked against numbers from other codes.
A community which supports this kind of data sharing is extremely important for reliable scientific reproducibility.

{\bf Reproducibility}--- Reproducibility of experimental results is one of the most important principles in the scientific community.
Numerical experiments should be held to as high standards, but often this is more difficult than it seems as code can change rapidly over time.
This is even more problematic for Monte Carlo algorithms where newly introduced features can alter the Markov chain resulting in slightly different numerical answers.
Furthermore, complex calculations rely upon an existing set of input and checkpoint files and produce similar numbers of files as output, making data provenance complicated.
As a simple measure to overcome this we output the input options and the git commit hash to the main output file and a UUID specific to the calculation in all output files which enables us and any other user to reproduce the results of a particular calculation.
We are fans of the IPython Notebook\cite{ipython} for data analysis as a way of storing the analysis and output together.  These notebooks also represent useful training aids.

{\bf Modern Standards}--- Languages continue to evolve and exploiting new developments can be a powerful tool in making code more flexible, portable and maintainable.  For example, the C interoperability features in Fortran 2003 make it much easier to combine existing code written in either language and so reduces the need to `reinvent the wheel'. One word of caution: new language features are implemented at different rates across different compilers, which are updated infrequently in some environments.  It is important to balance using new language features and staying away from the bleeding edge.  Regular testing against a variety of common compilers is vital in maintaining the portability of the code.

{\bf Bug fixing}--- Bug fixing in an academic environment is somewhat fraught given the inherently fluctional development community.  Whilst we have found the many bugs are prevented (or rather, discovered at time-of-creation) by code review, inevitably bugs remain to be discovered at a later time.
Whilst debugging is a universally hard problem, especially (as is often the case in academia) when the original student or researcher has moved on, we have found the approaches we discussed above crucial in mitigating this factor.  Good documentation, commenting and tests provide an indication of what the code should do (or at least what its author thought it should do!) and remove one layer of mystery.
We have also found code review an excellent strategy to aid this; having multiple developers review and understand a section of the codebase (albeit perhaps not on the same level as its author) aids the spread of knowledge throughout the development team and helps make it more likely that at least one person is capable of fixing the bug relatively quickly.
Once a bug is reported, it is triaged and a fix is proposed.  Following our standard code review process, it is then merged into the stable branch.
It is then important to update the test suite so that the bug remains fixed.
Who does this work can be problematic, especially in cases where the original is no longer working on HANDE.
Sometimes a code developer tracks the problem down.  In other cases we find the open source adage of `scratching your own itch' useful: the user who wants a bug fixed will (hopefully!) be suitably motivated to also fix it, given support and guidance from the wider development team.
We have found that this can be a powerful tool for encouraging users to become developers.

\section*{TRAINING}

The challenges facing someone joining a computational science project are multi-faceted: one must be knowledgeable in broad technical issues, the programming language(s) used as well as the theory of the underlying science.  
However, in practice, applied computer science is often attempted in academia without formal training. 
This requires that students learn on-the-job, but students often come highly motivated to learn new skills from day one.  Fortunately there are now excellent and affordable courses aimed at improving technical skills of computational scientists run by universities, national bodies (e.g.\ ARCHER in the UK\cite{archer}) and international groups.  We especially praise the impact of Software Carpentry\cite{software-carpentry}.

{\bf Introduction to HANDE}---
Ideally, the instruction given should be: `checkout the code and play around with it' and that should be sufficient; we aim for this to be the case.
New developers frequently comment that strategies mentioned in the previous section greatly help them in coming to grips with the code and in keeping initial motivation high.
We note this is a constant battle: additional features, optimisation and poor habits can cause the barrier of entry to creep up over time.
However, we find a mindful approach beneficial.
We recognise that initial impressions matter and so aim to make things as smooth as possible.  We find that the speed at which new developers learn is helped by
\begin{inparaenum}[\itshape a\upshape)]
\item a curated list of resources that cover the minimal amount of technical and scientific knowledge initially required;
\item writing a `toy' standalone code relevant to the problem (we get everyone to write a minimal FCIQMC program; another example is Ref.~\onlinecite{crawford_qc});
\item an introductory project which is both accessible and has a high chance of success, both technically and also as an appreciated contribution to the community.
\end{inparaenum}

Our experience is that highly-motivated students on moving away from the community willingly stay involved and enjoy doing so; this sets good examples for incoming students.
Informal, nonhierarchical, peer-based management greatly enhances this effect; learning happens organically in an environment where asking questions is easy and group discussion common.

{\bf Converting users to developers}--- By the very nature of academia, the development community around research software fluctuates.  Converting users into developers helps substantially in making a project sustainable, especially in niche fields.  In addition to attempting to minimise the barrier to entry, we find a powerful technique is to encourage users to `scratch their own itch': when a user has a feature request, we try to help them to implement it themselves (even if this takes more time than a core developer doing it themselves).  The time investment is typically rewarded surprisingly quickly.

{\bf Coding retreats}--- Engendering a development community and sharing knowledge across a geographically dispersed network is hard.  To this end we recently held a residential coding retreat.
Those in attendance were encouraged to implement a simple feature (\emph{i.e.}\ could be completed in the time available) of interest; coding review happened on-site.
We found this to be a good community-building format.  An important feature was to set aside substantial amounts of time for informal presentations and discussions, which provided a forum to discuss ongoing research as well as the codebase.

\section*{DISCUSSION}

We conclude with some examples of where our approach succeeded and where it failed, followed by an outlook on the future.

The development of a flexible, modular code supported by a training regime for new team members might appear to be a bet which may or may not pay off.  Our experiences show that it does pay off; in fact many of the approaches we discussed above were suggested naturally and adopted due to frustration with inefficiencies from \emph{not} doing them.  The impact on our work has been tremendous.  For example, two undergraduate students in a few months were able to propose, implement and test a new finite-temperature Monte Carlo approach in electronic structure\cite{blunt_density-matrix_2014}.  This would not have been possible if they had to start from scratch or from a monolithic, inpenetrable codebase.  Internal peer review has made our code more robust: review of recent improvements to the coupled cluster Monte Carlo\cite{thom_ccmc_2010} revealed a subtle bias when MPI parallelisation was used.
We have also found the community aspect in development to be important and have some unexpected benefits.  Recently several of us realised we were all struggling with a similar limitation in the code base and, as a result, embarked jointly on the (thankless) task of re-engineering some core data structures to provide additional flexibility.  It is unlikely this work would have taken place if everyone was instead just focussing on their own research project in isolation (which discourages this kind of improvement/tidying/maintance that benefits everyone) but doing so will actually open up new possibilities for all of us.

In other instances, we have been less successful.  One project on improving parallel scaling ended up running for almost a year, completely separate from the rest of the development.  Combining this with other work was painful: such large sets of changes are hard to review adequately and the resultant merge had lots of conflicts which had to be resolved manually.  We should have instead broken this work up into smaller sections rather than aiming for perfection in the first instance: our development model is better suited to continual refinement and incremental steps than large, radical changes.
Another example is from legacy work: a seemingly innocuous (largely stylistic) change three years ago introduced a bug in an extreme corner case which, naturally, was eventually triggered.  The problematic code dated back to before we systematically performed code reviews.  The developer who found the bug was able to spot it quickly in the affected procedure, but tracking it down to that point from some unusual results in production calculations was \emph{much} harder.
The last two cases are not where our development approach failed \emph{per se}, but rather where we failed it.
Whilst there is always the temptation to follow the `easy' course in the short term, in our experience this turns out to lead to pain later on---and often more quickly than anticipated!

As a project such as HANDE grows, there will be an increasing number of challenges in managing both the means of communication among the community as well as the direction of the project itself.
To ensure community growth, it is vital that the low barrier of entry be maintained, and one way we are planning to ensure this is to include developer tutorials which provide a step-by-step introduction to both the code \emph{and} our development practices.
Requiring novitiates to work through these tutorials has the three-fold goal of indoctrination into the coding and development standards, learning the structures of the project, and keeping the tutorials up-to-date themselves.
Often such tutorials are created on an \textit{ad hoc} basis, but such practices are to be encouraged so as to sustain the accessibility to all.  Indeed, the creation of tutorials aimed at users and developers would be a good introductory project when coupled with peer review.

We end with emphasising the benefits of an open source, collaborative approach, which we wholeheartedly endorse to the wider community.
A code which is well written and easily understandable makes it easier to spot mistakes, which can then be fixed quickly and results produced with an open source implementation can be reproduced with no ambiguity.
This enables scientists to spend more time pursing new ideas and less time resolving problems already solved by other groups, hence reducing the \emph{collective} time to productive science. 

{\bf Acknowledgments}---%
JSS and WMCF acknowledge the Thomas Young Centre under Grant No.~TYC-101.  WAV is grateful to EPSRC for a studentship, FDM for an Imperial College PhD scholarship, NSB to Trinity College, Cambridge for an External Research Studentship, JJS to the Royal Commission for the Exhibition of 1851 for a Research Fellowship and AJWT to the Royal Society for a University Research Fellowship. 
We acknoledge the Imperial College High Performance Computing Service and ARCHER via a RAP award and via the Materials Chemistry Consortium (Grant No.~EP/L000202).

\bibliographystyle{vancouver}
\bibliography{hande_wssspe2}

\end{document}